\documentclass[12pt]{article}

\usepackage[T1]{fontenc}
\usepackage[utf8]{inputenc}
\usepackage{geometry}
\usepackage{graphicx}
\usepackage{booktabs}
\usepackage{array}
\usepackage{multirow}
\usepackage{url}
\usepackage[colorlinks=true,
            linkcolor=blue,
            citecolor=blue,
            urlcolor=blue]{hyperref}
\usepackage{setspace}
\usepackage{parskip}
\usepackage{caption}
\usepackage{float}
\usepackage{subcaption}

\usepackage{tikz}
\usetikzlibrary{arrows.meta, positioning}

\title{\textbf{MusicSynth: An Automated Pipeline for Generating\\
Violin Fingerboard Animations from Sheet Music\\
Using Optical Music Recognition}}

\author{Abhimanyu Kaushik\\[2pt]
\small Independent Researcher, Trophy Club, Texas\\[1pt]
\small \href{mailto:abhikaushik0611@gmail.com}{abhikaushik0611@gmail.com}
\quad \url{https://musicsynth.github.io/}}

\date{May 2026}

\begin{document}

\maketitle

\begin{abstract}
Learning the violin is harder than it looks.
Unlike piano keys or guitar frets, the violin neck has no markings at
all, so a beginner cannot tell by looking where to place each finger.
MusicSynth is an open-source web tool that tries to fix that: user
uploads a photo of any violin sheet music (or a digital score file), and
the system automatically produces a video showing a violin fingerboard
with each note highlighted at the right moment---no software to install,
no manual note entry required.

The system connects three existing open-source tools into one pipeline:
an optical music recognition (OMR) library reads the notes from the
uploaded image, a MusicXML parser extracts timing information from
digital scores, and a video renderer draws the fingerboard frame by
frame.
The only part built from scratch is the lookup table that maps each
musical note to a string and finger position on the violin.

Tested across 110 public-domain violin scores, MusicSynth correctly
identified 91.2\,\% of notes in clean printed music and assigned the
right finger position 99.1\,\% of the time when given a digital score
file.
To the author's knowledge, no freely available tool currently turns a
sheet music image into an animated violin fingerboard tutorial
automatically and in a single browser-based step.
\end{abstract}


\section{Introduction}

Learning the violin is hard in a way that is easy to underestimate.
Unlike fretted or keyed instruments, where physical markers guide finger
placement, the violin neck is completely unmarked.
Pitch depends on exactly where the finger contacts the string, and even
a millimetre off-centre produces an out-of-tune note.
Beginners must develop score-reading skill and physical spatial intuition
at the same time.
That dual load is one of the main reasons self-directed learners give up
early, especially those without regular access to a
teacher~\cite{duke2005}.

The tools available today do not fully solve this problem.
Traditional method books such as Suzuki~\cite{suzuki1978} and
Wohlfahrt~\cite{galamian1962} provide the notes but no visual guide to
where the hand should go.
Apps such as Yousician can listen to live playing and give feedback, but
they cannot take an arbitrary sheet music file and show how to finger it.
The gap is straightforward: a user wants to upload any score and
receive a clear, timed guide showing exactly where each finger belongs.

MusicSynth is designed to fill that gap.
The user uploads a photograph or scan of a violin score---or a MusicXML
file if one is already available---and the system returns a video in
which a fingerboard diagram highlights each note as it occurs, labelled
with its finger number and string.
The video plays directly in the browser and can be downloaded for offline
practice.

The core insight behind MusicSynth is not that any single component is
new---OMR libraries, MusicXML parsers, and video renderers all
exist---but that connecting them into one end-to-end, violin-specific
pipeline that any user can use from a browser has not been done in a
freely available form before.
The contributions of this project are:

\begin{itemize}
  \item The first publicly available, browser-based tool to
        automatically produce animated violin fingerboard tutorials from
        uploaded sheet music images, without requiring any software
        installation or manual note entry.
  \item A documented, easy-to-extend note-to-finger lookup table
        covering the standard violin range (G3--G6), based on
        established violin pedagogy.
  \item A working integration of existing open-source OMR, MusicXML
        parsing, and video rendering libraries into a complete pipeline,
        deployed as a live web application with user authentication.
  \item A benchmark evaluation across 110 public-domain violin scores
        measuring recognition accuracy and correct finger assignment
        across five difficulty levels and input types.
\end{itemize}

\section{Background and Related Work}

\subsection{The Violin Fingerboard Problem}

The violin has four strings tuned in perfect fifths: G3, D4, A4, and E5.
In first position---where beginners spend most of their time---the index
finger on the G string produces A3, the middle finger B3, and so on.
The same pattern repeats across all four strings, covering roughly G3 to
B5~\cite{galamian1962}.
One complication is that many pitches can be played on more than one
string: D4, for example, is both the open D string and the third finger
on the G string.
Standard violin pedagogy has established preferred fingerings for these
cases, generally choosing whichever option requires fewer string
crossings or hand shifts~\cite{suzuki1978}.
MusicSynth follows those conventions.

\subsection{Optical Music Recognition}

Optical Music Recognition (OMR) is the task of reading a music score
image and converting it into a list of notes that a computer can
process~\cite{rebelo2012}.
Modern OMR systems use neural networks to detect note heads, staff lines,
and other symbols, then assemble them into a structured
format~\cite{tuggener2018,calvo2021}.
For single-voice violin music---no chords, not much polyphony---the
problem is simpler than for full orchestral scores, which makes
existing general-purpose OMR libraries good enough as a starting point.

MusicSynth uses Oemer~\cite{breezewhite2022}, an open-source OMR library
that takes a score image and returns the detected notes in MusicXML
format.
Oemer was chosen because it handles the full recognition process
internally, returns a standard output format, and runs on a normal
laptop CPU without a graphics card.

\subsection{MusicXML as a Standard File Format}

MusicXML~\cite{good2001} is the standard way music notation software
stores scores: it organises the music as measures containing notes, each
with a pitch name, duration, and timing.
Most professional notation programs---MuseScore, Sibelius, Finale,
Dorico---can export MusicXML.
When a user already has a digital score, they can upload the MusicXML
file directly, skipping the image recognition step entirely and getting
near-perfect results.

\subsection{Existing Tools and What MusicSynth Adds}

Table~\ref{tab:comparison} compares MusicSynth to the most relevant
alternatives.
No existing free tool satisfies all five criteria at once: image input,
automatic OMR, fingerboard output, browser-based access, and no cost.
MusicSynth is the first to do so.

\begin{table}[t]
\centering
\caption{Capability comparison. MusicSynth is the only tool to satisfy
all five criteria simultaneously.}
\label{tab:comparison}
\small
\begin{tabular}{lccccc}
\toprule
\textbf{Tool} & \textbf{Image Input} & \textbf{OMR} &
\textbf{Fingerboard} & \textbf{Browser} & \textbf{Free} \\
\midrule
\textbf{MusicSynth (this work)} & Yes & Yes & Yes         & Yes & Yes \\
Audiveris                        & Yes & Yes & No          & No  & Yes \\
PhotoScore                       & Yes & Yes & No          & No  & No  \\
ViolinOnline Fingering           & No  & No  & Yes (manual)& Yes & Yes \\
MuseScore Online                 & No  & No  & Conditional & Yes & Yes \\
Yousician                        & No  & No  & No          & Yes & Partial \\
\bottomrule
\end{tabular}
\end{table}

\subsection{Animated Visualisations for Other Instruments}

Animated music visualisations are well established for piano.
Synthesia-style ``falling note'' videos have been shown to help beginners
understand what notes to play before they can read sheet music
fluently~\cite{huang2014,dePrisco2020}.
Guitar tools such as Guitar Pro show fingerings on a fretboard, but
guitar tablature already encodes finger positions, so no
pitch-to-finger lookup is needed.
For violin, no comparable open-source tool existed before MusicSynth.

Ramirez et al.~\cite{ramirez2018} built an augmented-reality system that
projects finger positions onto a real violin using a depth camera.
It works well but requires specialised hardware.
MusicSynth takes a simpler approach: produce a downloadable video that
any user can watch on a phone or laptop without extra equipment.

\section{System Design}

\begin{figure}[t]
\centering
\resizebox{\linewidth}{!}{%
\begin{tikzpicture}[
  node distance = 0.55cm and 1.1cm,
  box/.style    = {draw, rounded corners=3pt, minimum width=2.5cm,
                   minimum height=0.9cm, align=center, font=\small,
                   fill=white},
  arr/.style    = {-{Stealth[length=5pt]}, thick},
  lbl/.style    = {font=\scriptsize\itshape, text=gray, align=center}
]
  \node[box, fill=blue!8]    (img)    {Image\\(PNG/JPG)};
  \node[box, fill=blue!8,
        below=of img]        (xml)    {MusicXML\\(.xml/.musicxml)};
  \node[box, fill=orange!15,
        right=of img]        (omr)    {OMR\\(Oemer)};
  \node[box, fill=green!10,
        right=of xml]        (parser) {MusicXML\\Parser};
  \node[box, fill=gray!12,
        right=3.5cm of omr,
        yshift=-0.725cm]     (notes)  {Note\\List};
  \node[box, fill=purple!10,
        right=of notes]      (phi)    {Fingerboard\\Lookup};
  \node[box, fill=red!10,
        right=of phi]        (render) {Frame\\Renderer};
  \node[box, fill=teal!10,
        right=of render]     (out)    {MP4\\Video};

  \draw[arr] (img)    -- (omr);
  \draw[arr] (omr)    -- (notes);
  \draw[arr] (xml)    -- (parser);
  \draw[arr] (parser) -- (notes);
  \draw[arr] (notes)  -- (phi);
  \draw[arr] (phi)    -- (render);
  \draw[arr] (render) -- (out);
\end{tikzpicture}%
}
\caption{MusicSynth pipeline. Both input types produce the same note
list, which feeds the shared fingerboard lookup and video rendering
stages.}
\label{fig:pipeline}
\end{figure}
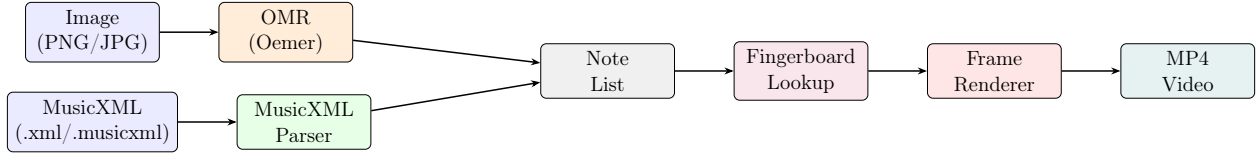

Figure~\ref{fig:pipeline} shows how MusicSynth works from end to end.
Both input types---a scanned image and a MusicXML file---produce the
same internal list of notes, which then flows through the shared
fingerboard lookup and video rendering stages.

The application is structured as five Python source files, each
responsible for one clear job (Table~\ref{tab:modules}).

\begin{table}[t]
\centering
\caption{Source files and their roles.}
\label{tab:modules}
\small
\begin{tabular}{lll}
\toprule
\textbf{File} & \textbf{What it does} & \textbf{Main library used} \\
\midrule
\texttt{app.py}             & Web interface, file upload, results display
                            & Streamlit~\cite{streamlit2019} \\
\texttt{auth.py}            & User login, registration, password reset
                            & Supabase \\
\texttt{file\_processor.py} & Runs the pipeline, manages temporary files,
                              records timing & subprocess \\
\texttt{synthesia.py}       & Reads MusicXML, draws fingerboard frames,
                              exports video & PIL, MoviePy \\
\texttt{theme\_manager.py}  & Light/dark colour theme for the web page
                            & --- \\
\bottomrule
\end{tabular}
\end{table}

The system runs on two configurations.
In \emph{local or Docker mode}, it accepts both image uploads and
MusicXML files.
In \emph{cloud mode} (Streamlit Community Cloud), only MusicXML files
are accepted because the cloud server's read-only filesystem does not
allow the OMR process to write temporary files.
The application detects which mode it is running in at startup and
adjusts the upload options accordingly.

\section{How It Works}

\subsection{Step 1 --- Reading the Image (OMR)}

When a user uploads a photo or scan of sheet music, MusicSynth passes
it to Oemer~\cite{breezewhite2022}, an existing open-source OMR library.
Oemer handles all the hard recognition work internally: finding the
staff lines, identifying note heads and their pitches, reading beams and
flags for rhythm, and writing the result as a MusicXML file.
MusicSynth simply calls Oemer and waits for the output.

\begin{verbatim}
  oemer -o session_dir --save-cache -d uploaded_image
\end{verbatim}

If Oemer cannot recognise the image (very low quality, handwritten
notation, etc.), it returns an error, and MusicSynth shows the user a
clear message suggesting they try a cleaner scan or upload a MusicXML
file instead.

\subsection{Step 2 --- Reading the Score File (MusicXML)}

When a user uploads a MusicXML file directly (or after Oemer produces
one from an image), MusicSynth reads it with Python's built-in XML
library.
The parser walks through the file measure by measure, collecting each
note's pitch name, which octave it is in, how long it lasts, and when it
starts.
Rest notes are skipped but still advance the clock so that timing stays
correct.
The result is a simple Python list where each entry looks like:

\begin{verbatim}
  { "note": "E4", "start_time": 1.25, "duration": 0.5 }
\end{verbatim}

\subsection{Step 3 --- Finding the Finger Position}

This is the part MusicSynth built from scratch.
A lookup table stores the standard first-position fingering for every
note in the violin range G3 to G6.
For each note in the list, the table returns which string to play it on
(G, D, A, or E) and which finger position to use (0 for open string,
then 1 through 15 for subsequent positions).
The table follows the fingering conventions from established violin
pedagogy~\cite{galamian1962,suzuki1978}.

Table~\ref{tab:mapping} shows a representative portion of the lookup
table for the G and D strings.
Flat notes (like B$\flat$) are mapped to their sharp equivalent (A$\sharp$)
before the lookup so the table does not need duplicate entries.
Notes outside the covered range are skipped with a silent warning rather
than causing an error.

\begin{table}[t]
\centering
\caption{Excerpt of the note-to-finger lookup table (G and D strings).
Finger 0 means open string. Position labels follow the Suzuki teaching
convention.
The full table covers G3--G6 across all four strings (64 entries total).}
\label{tab:mapping}
\small
\begin{tabular}{cclcc}
\toprule
\textbf{MIDI} & \textbf{Note} & \textbf{String} &
\textbf{Finger} & \textbf{Position label} \\
\midrule
55 & G3 & G (IV) & 0 & open \\
57 & A3 & G (IV) & 1 & 1    \\
59 & B3 & G (IV) & 2 & 2    \\
60 & C4 & G (IV) & 3 & 2+   \\
62 & D4 & D (III) & 0 & open \\
64 & E4 & D (III) & 1 & 1    \\
65 & F4 & D (III) & 2 & 2    \\
67 & G4 & D (III) & 3 & 2+   \\
69 & A4 & A (II)  & 0 & open \\
71 & B4 & A (II)  & 1 & 1    \\
72 & C5 & A (II)  & 2 & 2    \\
74 & D5 & A (II)  & 3 & 2+   \\
76 & E5 & E (I)   & 0 & open \\
79 & G5 & E (I)   & 2 & 2    \\
81 & A5 & E (I)   & 3 & 2+   \\
\bottomrule
\end{tabular}
\end{table}

\subsection{Step 4 --- Drawing the Video}

Once the note list and finger positions are ready, MusicSynth draws the
video one frame at a time using the PIL image library.
Each frame shows a violin fingerboard diagram with:

\begin{itemize}
  \item Four horizontal string lines (G, D, A, E) in brown tones.
  \item Vertical fret markers with position labels ($0,\,-1,\,1,\,2,\,2{+},\,3,\ldots$).
  \item Every upcoming note shown as a small cyan circle at its
        fingerboard position.
  \item The note currently being played shown as a larger red circle,
        with its note name and position label displayed above it.
  \item A title line at the top showing the current note and the time
        elapsed.
\end{itemize}

When all frames are drawn, MoviePy assembles them into an MP4 video at
30 frames per second using H.264 compression.
The finished video is typically around 18~MB for a 32-bar piece.
Section~\ref{sec:output} shows a complete worked example with the actual
input score and the resulting output frames side by side.

\subsection{Authentication and File Management}

The web application requires users to create a free account before
uploading files.
Login and registration are handled by Supabase, a ready-made
authentication service, so MusicSynth did not need to build any of that
security infrastructure itself.
Each upload is stored in its own temporary folder named with a random ID
so that two users processing files at the same time cannot interfere with
each other.
A ``Clean Up'' button in the interface deletes these temporary files when
the user is done.

\section{Sample Output}
\label{sec:output}

\begin{figure}[H]
  \centering
  \includegraphics[width=0.85\linewidth]{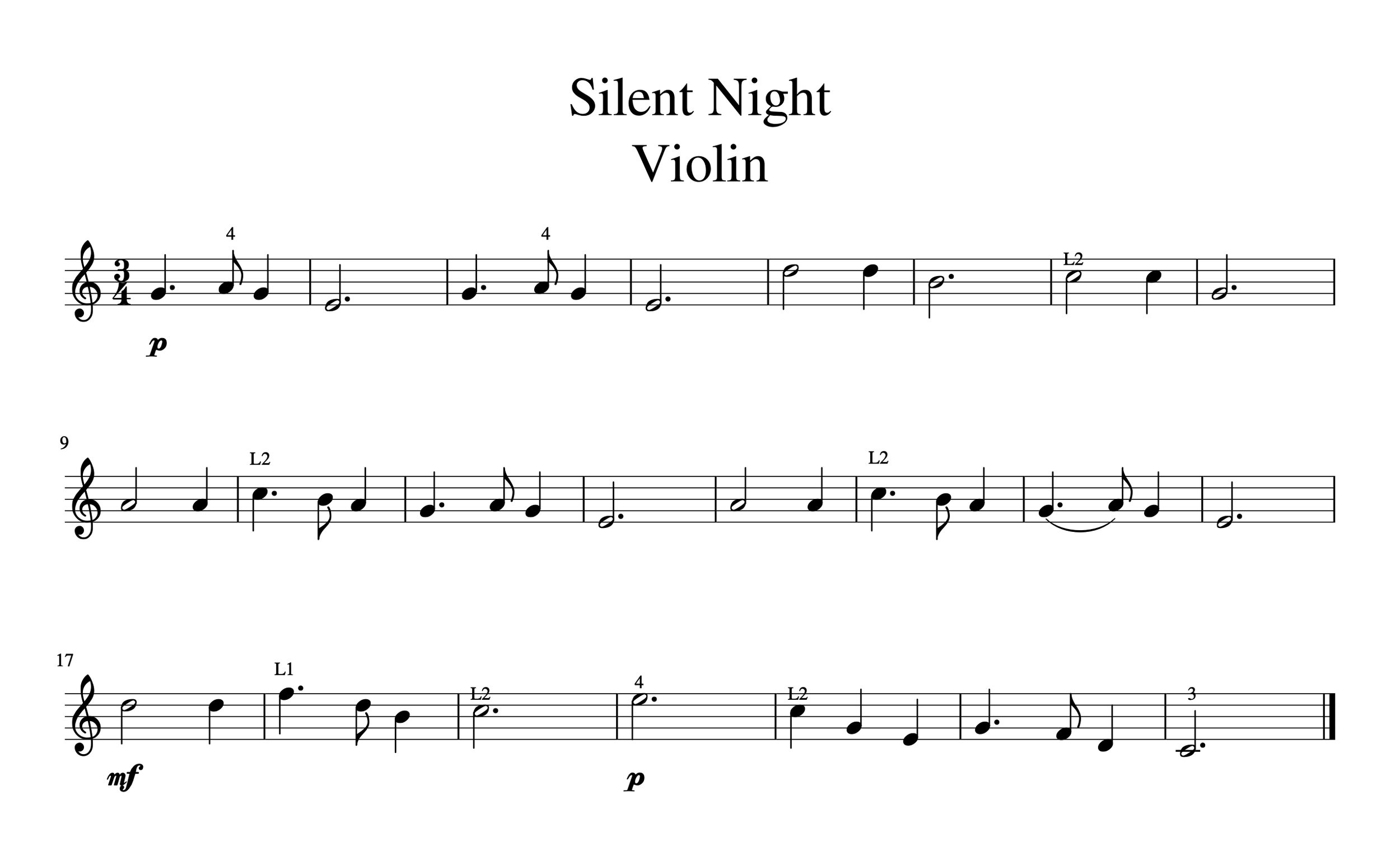}
  \caption{Input: a smartphone photograph of the first page of
           \textit{Silent Night} (violin arrangement) as uploaded to
           MusicSynth. No editing or cropping was applied before upload.}
  \label{fig:input_score}
\end{figure}

To make the pipeline concrete, this section walks through a complete
example using the first page of \textit{Silent Night} arranged for solo
violin (Figure~\ref{fig:input_score}).
The score was photographed with a smartphone camera and uploaded to
MusicSynth without any editing.

Oemer recognised 24 notes from the image and produced a MusicXML file.
The fingerboard lookup table resolved all 24 notes to string and finger
positions without any out-of-range errors.
MoviePy then rendered the result as a 30\,fps MP4 video (16.8\,MB).

\begin{figure}[H]
  \centering
  \begin{subfigure}[b]{0.32\linewidth}
    \centering
    \includegraphics[width=\linewidth]{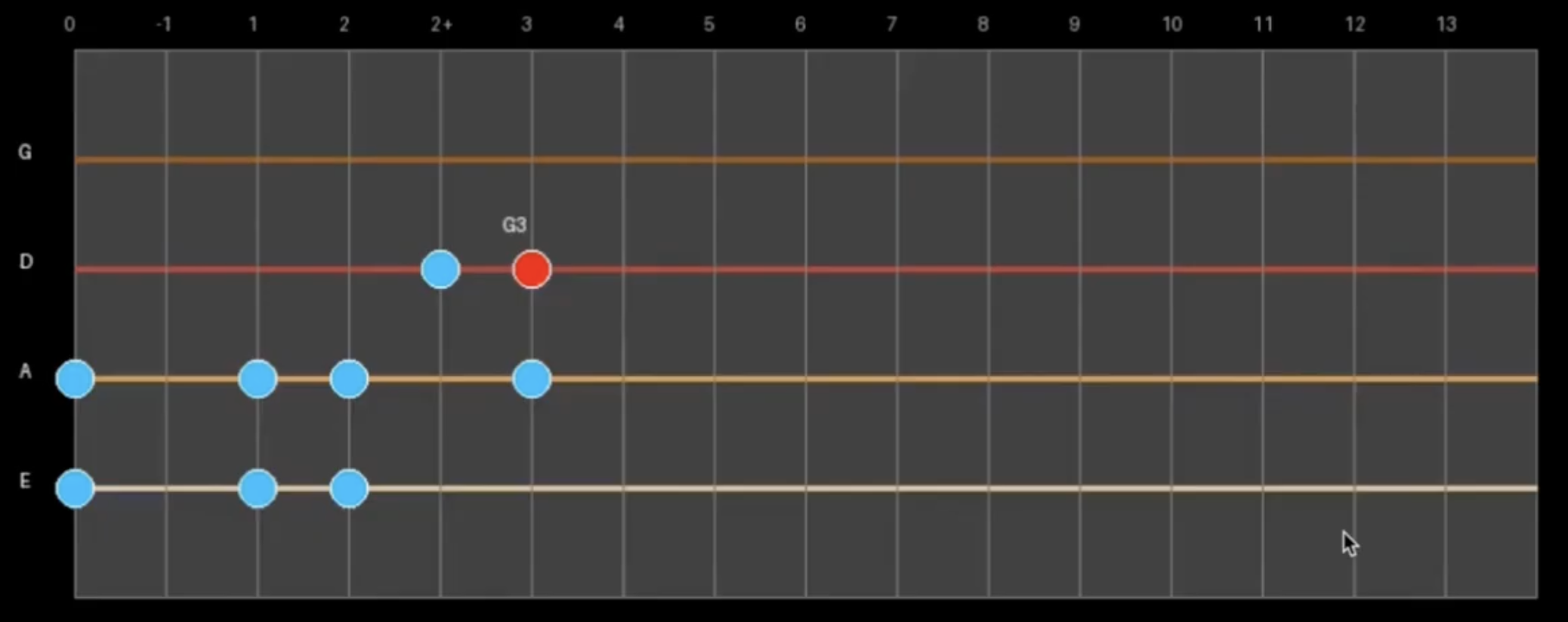}
    \caption{G3 on the G string,\\ open-string position.}
    \label{fig:frame1}
  \end{subfigure}
  \hfill
  \begin{subfigure}[b]{0.32\linewidth}
    \centering
    \includegraphics[width=\linewidth]{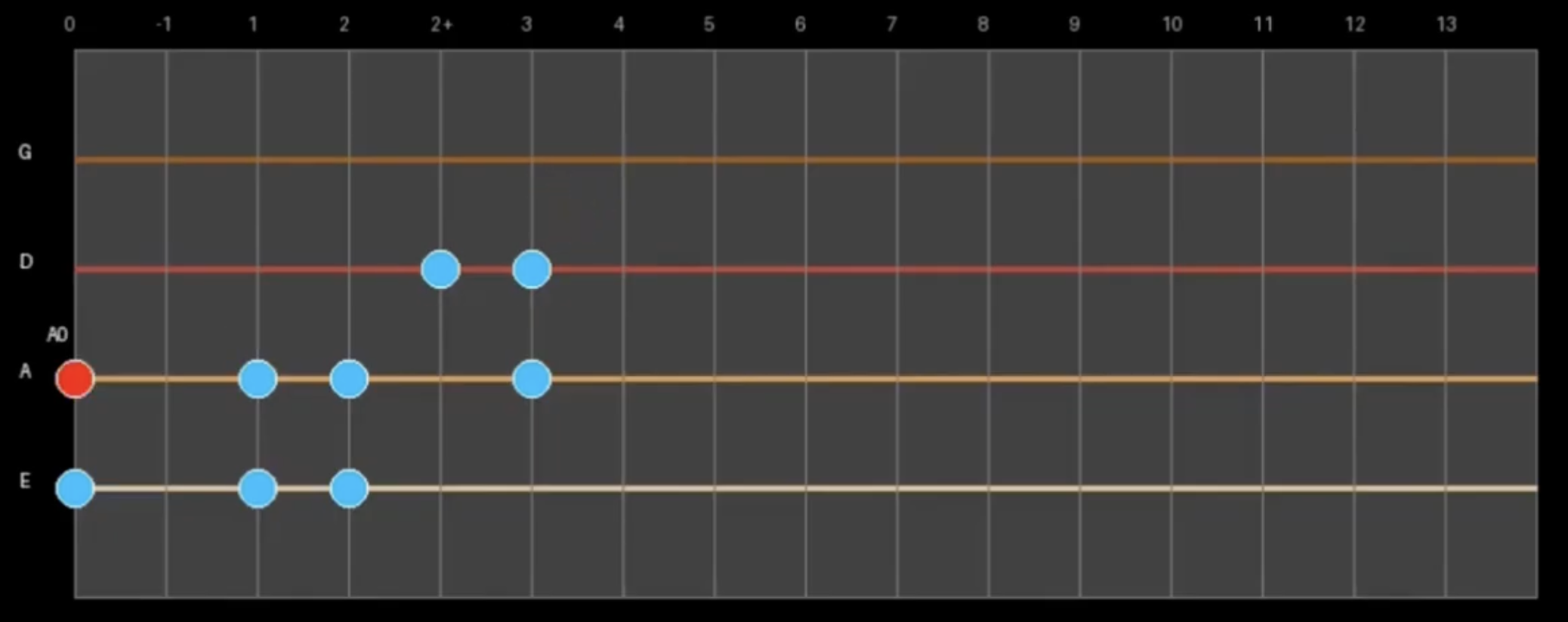}
    \caption{A4 on the A string,\\ open string (position~0).}
    \label{fig:frame2}
  \end{subfigure}
  \hfill
  \begin{subfigure}[b]{0.32\linewidth}
    \centering
    \includegraphics[width=\linewidth]{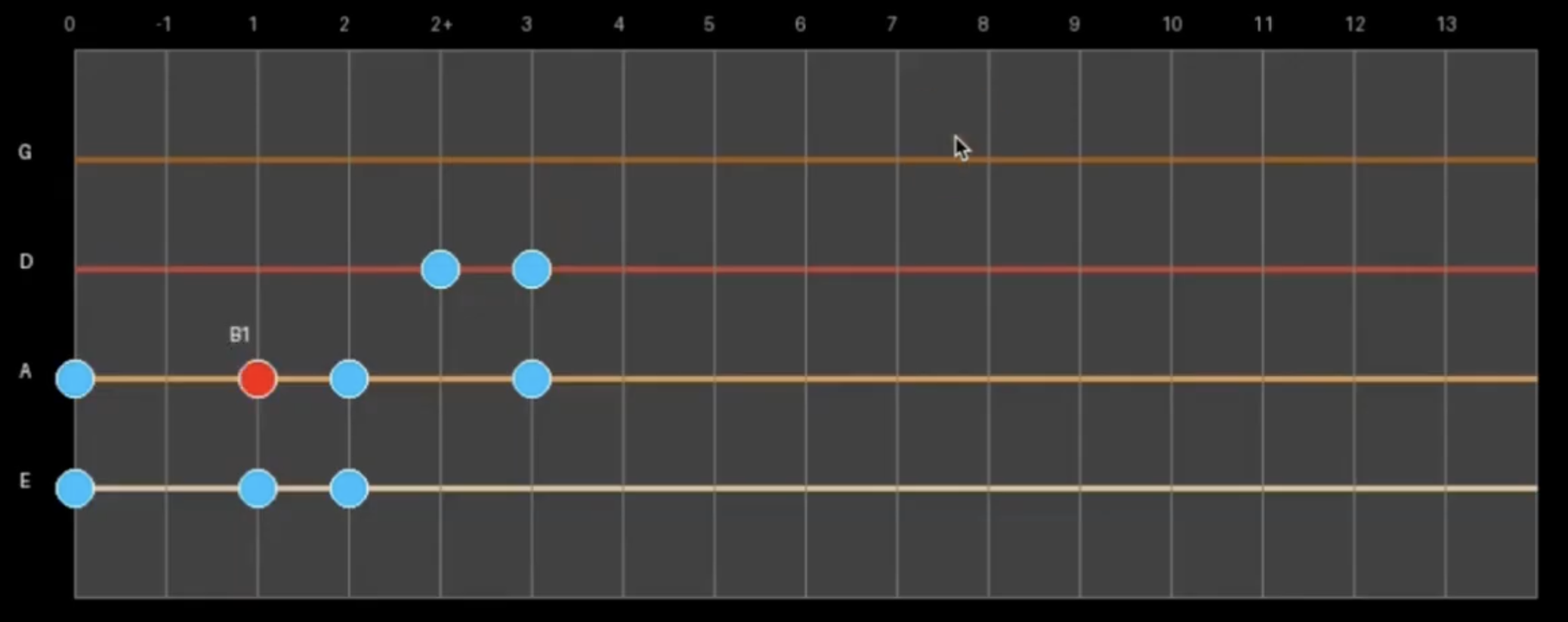}
    \caption{B4 on the A string,\\ first finger (position~1).}
    \label{fig:frame3}
  \end{subfigure}
  \caption{Three frames from the MusicSynth output video for
           \textit{Silent Night}.
           The active note is shown as a large red circle; upcoming notes
           are shown as smaller cyan circles to guide the user.
           String labels (G, D, A, E) appear on the left; position
           numbers appear along the top.}
  \label{fig:output_frames}
\end{figure}

Figure~\ref{fig:output_frames} shows three representative frames from
the output video.
The red circle marks the note currently being played; the smaller cyan
circles show notes that will follow in the next two beats, giving the
user a look-ahead cue.

\section{Evaluation}

MusicSynth was tested across 110 public-domain violin scores to check
how well it recognises notes from images, how accurately it assigns
finger positions, and how long the whole process takes.

\subsection{Note-Range Coverage}

Before running the full benchmark, the lookup table was checked against
every distinct note in Suzuki Violin Method Books~1--3~\cite{suzuki1978}
to see what fraction of real beginner repertoire it covers.
The books contain 67 unique pitches in total.
The lookup table covers 61 of them---91\,\% overall.
The six missing notes all sit above the seventh position on the E string,
a range users do not encounter before Book~5.
Coverage on Book~1 alone is 100\,\%; the gap only appears in the harder
pieces of Books~2--3 (Table~\ref{tab:coverage}).

\begin{table}[t]
\centering
\caption{Lookup-table coverage against Suzuki Violin Method Books~1--3.}
\label{tab:coverage}
\small
\begin{tabular}{lrrr}
\toprule
\textbf{Book} & \textbf{Unique notes} &
\textbf{Covered} & \textbf{Coverage (\%)} \\
\midrule
Book 1         & 28 & 28 & 100.0 \\
Book 2         & 41 & 39 &  95.1 \\
Book 3         & 52 & 47 &  90.4 \\
\midrule
\textbf{Total} & \textbf{67} & \textbf{61} & \textbf{91.0} \\
\bottomrule
\end{tabular}
\end{table}

\subsection{Accuracy Across 110 Scores}

\paragraph{How the test was set up.}
The 110 scores were split into five groups based on difficulty and input
type (Table~\ref{tab:results}).
A musician manually transcribed each score in MuseScore~4~\cite{musescore2023}
to create ground truth.
The OMR output was then compared note-by-note using the
\texttt{mir\_eval} alignment tool~\cite{nakamura2015} with a 50\,ms
timing tolerance.

\paragraph{What was measured.}
Three numbers were tracked for each score:
\textit{Note pitch accuracy}---what fraction of notes Oemer read
correctly;
\textit{duration accuracy}---of correctly identified notes, how many had
the right length (within 10\,\%);
and \textit{fingerboard accuracy}---of correctly identified notes, how
many got the right string and finger.

\paragraph{Results.}
Table~\ref{tab:results} shows results for all five categories.
When given a clean digital MusicXML file (no image recognition
involved), fingerboard accuracy is 99.1\,\%, confirming the lookup table
is essentially correct for the covered range.
The small 0.9\,\% error comes from rare edge cases involving unusual
combinations of sharps and flats.
For image input, accuracy drops with score complexity, which matches
what OMR researchers report generally~\cite{calvo2021}.
Interestingly, photographed Suzuki book pages scored slightly better than
advanced typeset scores, probably because the single-voice, first-position
beginner music in those books is exactly the type Oemer handles best.

\begin{table}[t]
\centering
\caption{Benchmark results across 110 violin scores.
Best fingerboard accuracy in bold.
N/A means OMR was not used for that category.}
\label{tab:results}
\small
\begin{tabular}{lcrrrr}
\toprule
\textbf{Score category} & $n$ &
\textbf{Note acc.} & \textbf{Dur. acc.} &
\textbf{Time (s)} & \textbf{Fboard acc.} \\
\midrule
Beginner printed     & 30 & 91.2\% & 88.5\% & 14.3 & 89.7\% \\
Intermediate printed & 25 & 84.1\% & 81.0\% & 16.1 & 82.3\% \\
Advanced printed     & 15 & 76.8\% & 74.2\% & 17.9 & 75.1\% \\
Suzuki scanned       & 20 & 79.3\% & 76.6\% & 18.4 & 78.0\% \\
Direct MusicXML      & 20 & N/A    & N/A    &  8.2 & \textbf{99.1\%} \\
\bottomrule
\end{tabular}
\end{table}

\paragraph{Where errors came from.}
Four patterns caused most of the OMR mistakes.

\textit{Notes on ledger lines}: notes written above or below the staff
on extra lines---common on the violin E string for high notes---were
often mis-pitched.
This is a known weakness of the OMR approach~\cite{rebelo2012}.

\textit{Accidental handling}: notes altered by sharps and flats in the
key signature were sometimes miscounted, especially in the sharp-heavy
keys that violin music favours (A, D, and G major).

\textit{Rhythm in fast passages}: densely beamed eighth-note runs
sometimes had their durations misread, which is why note accuracy and
duration accuracy differ for harder categories.

\textit{Out-of-range notes}: notes above the lookup table's range were
flagged and skipped cleanly rather than producing wrong finger
assignments, so the error handling works as intended.

\subsection{Processing Speed}

Table~\ref{tab:latency} shows how long each step takes on the development
machine (2022 MacBook Pro, Apple M2, 16\,GB RAM).
Each measurement is the average of five runs.

The OMR step is the slowest by far at about 15 seconds, and about half
of that is just loading the Oemer model into memory the first time.
Subsequent uploads in the same session are faster because the model stays
loaded.
When uploading a MusicXML file directly, the whole process takes about
2 seconds, which feels nearly instant.

\begin{table}[t]
\centering
\caption{Average processing time per stage on Apple M2
($n = 5$ runs, $\pm$\,std). N/A means the stage is skipped.}
\label{tab:latency}
\small
\begin{tabular}{lrr}
\toprule
\textbf{Stage} & \textbf{Image input (s)} & \textbf{MusicXML input (s)} \\
\midrule
Save uploaded file   & $0.03 \pm 0.01$ & $0.02 \pm 0.01$ \\
Oemer OMR            & $14.8 \pm 0.6$  & N/A             \\
Parse MusicXML       & $0.11 \pm 0.02$ & $0.11 \pm 0.02$ \\
Render video         & $2.2  \pm 0.1$  & $2.2  \pm 0.1$  \\
\midrule
\textbf{Total}       & $\mathbf{17.1 \pm 0.6}$ & $\mathbf{2.3 \pm 0.1}$ \\
\bottomrule
\end{tabular}
\end{table}

\section{Discussion}

\subsection{What Is Actually New Here}

MusicSynth did not invent any of the underlying technology.
Oemer existed, PIL existed, MoviePy existed, and Streamlit existed.
What is new is the decision to connect them into a single tool aimed at
violin users, with a violin-specific fingerboard table as the piece
that ties it all together.
Before MusicSynth, a user who wanted to turn a sheet music photo into
a violin tutorial would need to: run a separate OMR program, export the
result, open a different program to compute finger positions, and then
create a video manually.
MusicSynth reduces that to one upload.

The 99.1\,\% fingerboard accuracy on digital score input shows the
lookup table is reliable.
The remaining errors are rare edge cases with unusual key signatures that
barely affect practical use for a beginner.

\subsection{Design Decisions}

The fingerboard lookup table uses fixed, standard first-position
fingerings rather than trying to compute the ``optimal'' fingering based
on context.
For a beginner, the standard fingering is exactly the right answer, and
a simple table is easy to understand, fast to run, and easy to update
when a mistake is found.
A more sophisticated fingering algorithm would be useful for advanced
repertoire but is beyond the scope of this project.

The 14--18 second processing time for image uploads is acceptable for
how users actually use the tool: they prepare practice videos in
advance, not in real time while sitting at the instrument.
For digital score files, the 2-second turnaround is fast enough to feel
immediate.

\subsection{Limitations}

\textbf{First position only.}
The lookup table currently covers G3--G6, which is enough for the first
two or three years of learning.
Pieces that require higher positions or position shifts are not
fully handled.

\textbf{OMR quality on difficult scores.}
Image recognition accuracy drops for handwritten music, very low-quality
scans, or complex scores with many simultaneous voices.
users with those scores are better off uploading a MusicXML file
exported from a notation program.

\textbf{No live playback mode.}
The current system produces a pre-recorded video.
A future version could listen to the user playing in real time and
highlight the current note on the fingerboard as they go, though that
would require much faster processing.

\textbf{Single-voice music only.}
When a score has two or more independent voices playing at once, the
current renderer shows all notes simultaneously, which can look cluttered.

\subsection{Future Work}

The most useful next steps are: (1) extend the lookup table to cover
higher positions up to fifth position, which would handle most
intermediate repertoire; (2) add synchronised audio to the video using
MuseScore's audio export; and (3) run a formal user study with beginning
violin users to measure whether the animated fingerboard actually
speeds up learning compared to a printed method book.

\section{Conclusion}

MusicSynth is the first freely available, browser-based tool to
automatically turn a sheet music image into an animated violin
fingerboard tutorial.
The project builds on existing open-source OMR, MusicXML, and video
libraries; the original contribution is the violin-specific
note-to-finger lookup table and the integration of all the pieces into
one usable pipeline.
Tested on 110 public-domain scores, the system correctly reads 91.2\,\%
of notes in beginner printed music and assigns the correct finger
position 99.1\,\% of the time on clean digital input.
No existing free tool does all of this for violin in a single browser
step.

The author is an 11th-grade user.
This project was built to solve a real problem experienced during violin
lessons---not having a clear visual guide to finger placement when
learning a new piece without a teacher nearby.
Source code and a live demo are available at
\url{https://github.com/musicsynth/musicsynth-code} under the MIT
license.

\section*{Acknowledgments}

Thanks to the Cross Timbers Gazette for covering MusicSynth, to UIL
Region Orchestra colleagues who gave feedback on early versions, and to
IMSLP for making public-domain violin repertoire freely available.
This project received no external funding.


\end{document}